\documentclass[graybox]{svmult}

\usepackage{type1cm}        % activate if the above 3 fonts are
\usepackage{makeidx}         % allows index generation
\usepackage{graphicx}        % standard LaTeX graphics tool
\usepackage{multicol}        % used for the two-column index
\usepackage[bottom]{footmisc}% places footnotes at page bottom

\usepackage{newtxtext}       % 
\usepackage[varvw]{newtxmath}  

\usepackage{times}

\usepackage[numbers,sort&compress]{natbib}
\usepackage{graphicx}
\usepackage{amsmath,amscd,amsthm}
\usepackage{color}

\newcommand{\beq}{\begin{eqnarray}}
	\newcommand{\eeq}{\end{eqnarray}}
\def\be{\begin{equation}}
	\def\ee{\end{equation}}
\newcommand\eq[1]{Eq.~(\ref{#1})}

\makeindex             % used for the subject index
\begin{document}

\title*{ A Brief Survey of  Fluctuation-induced Interactions in Micro- and Nano-systems and One Exactly Solvable Model as Example}
% Use \titlerunning{Short Title} for an abbreviated version of
% your contribution title if the original one is too long
\author{Daniel Dantchev\orcidID{0000-0002-4762-617X} and \\ Nicholay Tonchev\orcidID{0000-0003-4088-7550}}
\authorrunning{Fluctuation-induced Interactions} 
%\authorrunning{Fluctuation-induced Forces}
\institute{Daniel Dantchev \at Institute of Mechanics, Bulgarian Academy of Sciences, Acad. G. Bontchev bl. 4, 1113 Sofia, Bulgaria, and Max-Planck-Institut f\"{u}r Intelligente Systeme, Heisenbergstrasse 3, D-70569 Stuttgart, Germany, \email{daniel@imbm.bas.bg}
\and  Nicholay Tonchev \at Institute of Solid State Physics, Bulgarian Academy of Sciences,1784 Sofia, Bulgaria, 
\email{nicholay.tonchev@gmail.com}}
%
% Use the package "url.sty" to avoid
% problems with special characters
% used in your e-mail or web address
%
\maketitle

\abstract{
	Fluctuations exist in any material object $A$. If  $A$ has non-zero temperature $T$, one speaks about thermal fluctuations. If  $A$ is at very low $T$,  the fluctuations are  of quantum origin. Interesting effects appear  if two bodies $A$ and $B$ are separated by a fluctuating medium $C$ (say a vacuum, or a fluid close to its {\it critical point}) when the fluctuations are long-ranged, i.e., they decay according to a power-law with the distance. Then the changes of fluctuations in $C$ due to the surfaces and constituents of $A$ are also felt by $B$, and \textit{vice versa},  which leads to a fluctuation induced force (FIF) between them. This force persists in addition to the direct influence of $A$ on $B$ (say, via gravity or Coulomb's force). These FIF's can be of  attractive  or repulsive  character. They may play  crucially important role on phenomena involving objects with length scale comparative with the Universe,  as well as to the tiny objects relevant for  MEMS and NEMS.  In the current article we present some basic facts for the FIF and their diversity. Then on the example of one dimensional Ising model with a defect bond we present some new analytical results for such forces.}

\section{Introduction in Fluctuation-induce Forces: A Brief Review}
\label{sec:introduction}
We consider two macroscopic or mesoscopic material bodies $A$ and $B$, separated by a fluctuating medium $C$. We always suppose that the degrees of freedom can enter and leave the region between the interacting
objects. There are, however, two important subcases - one in which the $C$ is in some contact with a reservoir, i.e., its constituents can enter and leave the part of the space occupied by $A$ and $B$. In this case on speaks about  the  \textit{Casimir force}. In the second case, the system is itself bounded, so that some quantity characterizing the amount of material in $C$ is conserved. Then, on speaks about the recently introduced, see the Letter  \cite{DR2022}, and still not well studied, \textit{Helmholtz force}. Both these forces are examples of the fluctuation induced forces and they do exist because the medium $C$ fluctuates and, when the force decays algebraically with the distance so do the correlations of the fluctuations in $C$. Many other examples of such forces are presented in Ref. \cite{DD2022}. 

Maybe the first study of a fluctuation induced force is due to A. Einstein \cite{E07}. Having in mind a standard capacitor at nonzero temperature $T$, as early as in 1907 he studied the  voltage fluctuations between its plates and even concluded that the corresponding effect can be measured.  

Presently, the most famous example of a fluctuation-induced interaction
is the \textit{quantum electrodynamic} (QED) Casimir effect \cite{C48,MT97,KG99,M2001,BKMM2009}. Nowadays, investigations devoted to that effect are performed on many fronts of research ranging from attempts to unify the four fundamental forces of nature \cite{MT97,M2001,M2004} to rather practical issues such as the design and the performance of MEMS and NEMS  \cite{GLR2008,KMM2011,RCJ2011,FAKA2014,FMRA2014}.

In the QED Casimir effect the medium $C$ is the vacuum; the presence of two conducting plates (the interacting objects $A$ and $B$) modifies the zero point energy of the
electromagnetic field and leads to an attractive force (normalized per area, i.e. to the so-called Casimir pressure)
 \begin{equation}\label{qCasimirf}
	F_{\rm Cas}^{(\rm QED)}(L)=-\frac{\pi^2}{240}\frac{\hbar c}{L^4} = -1.3
	\times 10^{-3}\frac{1}{(L/{\rm \mu m})^4} {\rm \frac{N}{m^2}},
\end{equation}
where $L$ is the separation between the plates, $\hbar$ and $c$ are the Planck constant ($\hbar=h/(2\pi)$), and the speed of light in vacuum, respectively. The QED Casimir effect is one of the rare manifestations of
quantum physics at the macroscopic scale, like super-conductivity and super-fluidity.  

At non-zero
temperature, as it shall be expected, the thermal fluctuations come into play, giving rise to additional temperature-dependent interactions. When applied to realistic materials, the material properties of the bodies $A, B$ and the medium $C$ get also involved via their  general dielectric and conductive properties. This has been done by	E. M. Lifshitz \textit{et al.}\cite{L56,L.E.Dzyaloshinskii1961}, see also \cite{MN76,P2006}. There the material properties enter  via the  frequency-dependent dielectric permittivities $\varepsilon^{(A)}(\omega)$,  $\varepsilon^{(B)}(\omega)$, and $\varepsilon^{(C)}(\omega)$. In the limit of small separations (but still large compared with molecular scales) the Casimir force approaches the more familiar van der Waals force \cite{DLP61r,BKMM2009}.  
From Lifshitz theory one  can infer
\cite{DLP61r} that there is
a possibility to observe QED Casimir
{\it repulsion} in the film geometry if the two  half-spaces (A) and (B) forming
the plates and confining the film (C) exhibit permittivities which fulfill the
relationship
\begin{equation}
	\varepsilon^{(B)}(i\xi)<\varepsilon^{(C)}(i\xi)<\varepsilon^{(A)}(i\xi).
	\label{repulsion_condition}
\end{equation}
Experimentally repulsion occurs if the inequality in \eq{repulsion_condition}
holds over a \textit{sufficiently wide
	frequency range}. Actually this is a widespread phenomenon shared by all
substrate-fluid systems which show complete wetting  \cite{Di88}. Accordingly
Casimir repulsion is a common feature and has been already observed - see, e.g., Ref. \cite{MCP2009}.

Thirty years after H. B. G. Casimir, in 1978 M. Fisher and P-G. de Gennes \cite{FG78} have shown that a very similar fluctuation-induced effect exists in fluids. This is now the widely-investigated \textit{critical Casimir effect} (CCE). It results from the fluctuations of an order parameter and, more generally, from the thermodynamics of the medium supporting that order parameter in the vicinity of a critical point. Recently, a review on the exact results available for the CCE has been published in Ref. \cite{DD2022}.  On different aspects of this effect overviews can be found in \cite{Krech1994,BDT2000,MD2018,DD2022,Gambassi2023}. For the critical Casimir effect (CCE) the expression, analogous to \eq{qCasimirf}, exist at the critical point $T=T_c$ of the fluid $C$. 
For the $(d=3)$-dimensional system one can write the critical Casimir force (CCF) per unit area, i.e., the Casimir pressure, in the form 
\begin{equation}\label{DimensionsFCas}
	F_{{\rm Cas}}^{(\tau)}(T=T_c, L)\simeq8.1\times10^{-3}
	\dfrac{\Delta^{(\tau)}(d=3)}{(L/\mu{\rm m})^{3}}\dfrac{T_{c}}{T_{\rm roon}}\dfrac{{\rm N}}{{\rm m^{2}}},
\end{equation}
where $T_{\rm room}=20$ $^\circ$C (293.15 K). Here $\Delta^{(\tau)}$ is the so-called \textit{Casimir amplitude} that depends on the bulk and surface \textit{universality classes} (see below) of the system and the applied \textit{boundary conditions} $\tau$. For most systems and boundary conditions one has $\Delta^{(\tau)}(d)={\cal O}(1)$.

Thus, the both forces, the quantum and the critical one, can be of the \textit{same} order of magnitude, i.e.,
they both can be essential, measurable and obviously significant at or below the micrometer length scale. Let us stress that $\Delta^{(\tau)}(d)$ can be both \textit{positive and negative}, i.e., $F_{{\rm Cas}}^{(\tau)}(T,L)$ can be both \textit{attractive and repulsive}. The accepted  terminology terms the negative force as attractive one.

In recent Letter  \cite{DR2022} we have introduced and studied a \textit{Helmholtz fluctuation induced force}. It is a force in which an integral quantity value of the order parameter characterizing the system is fixed (say the total magnetization in the system). We stress, that in standard envisaged applications of, say, the equilibrium Ising model to binary alloys or binary liquids, the case with order parameter fixed must be addressed, provided that one considers finite systems and insists on a rigorous analytical treatment. In Refs. \cite{DR2022,Dantchev2023b}  via deriving there exact results on the example of Ising chain with fixed magnetization under periodic and antiperiodic boundary conditions, we have established a very different behavior of the Helmholtz force from that one of the Casimir force, in the same model and under the same boundary conditions. It is interesting to note that, actually, under periodic boundary conditions the studied Helmholtz force has a behavior similar to the one appearing in some
versions of the big bang theory --- strong repulsion at high
temperatures, transitioning to moderate attraction for intermediate
values of the temperature, and then back to repulsion,
albeit much weaker than during the initial period of highest
temperature.  

In order to be concrete and avoid any misunderstandings, let us remind  the definitions of the critical Casimir and Helmholtz forces --- the Casimir force in the grand canonical ensemble (GCE), and its analogue in the canonical ensemble (CE) --- the  Helmholtz force . 

In the general case, we envisage a  $d$-dimensional system with a film geometry $\infty^{d-1}\times L$, $L\equiv L_\perp$, and with boundary conditions $\zeta$ imposed along the spatial direction of finite extent $L$.  Let us take ${\cal F}_{ {\rm tot}}^{(\zeta)}(L,T,h)$ to be the total (Gibbs) free energy of such a system within the GCE, where $T$ is the temperature and $h$ is the magnetic field. Then, if   $f^{(\zeta)}(T,h,L)\equiv \lim_{A\to\infty}{\cal F}_{ {\rm tot}}^{(\zeta)}/A$  is the free energy per area $A$ of the system, one can define the Casimir force for critical systems in the grand-canonical $(T-h)$-ensemble, see, e.g. Ref. \cite{Krech1994,BDT2000,MD2018,DD2022,Gambassi2023}, as: 
\begin{equation}
	\label{CasDef}
	\beta F_{\rm Cas}^{(\zeta)}(L,T,h)\equiv- \frac{\partial}{\partial L}f_{\rm ex}^{(\zeta)}(L,T,h)
\end{equation}
where
%\end{document}
\begin{equation}
	\label{excess_free_energy_definition}
	f_{\rm ex}^{(\zeta)}(L,T,h) \equiv f^{(\zeta)}(L,T,h)-L f_b(T,h)
\end{equation}
is the so-called excess (over the bulk) free energy per area and per $\beta^{-1}=k_B T$.

Along these lines, if $M$ is the fixed value of the total magnetization,  the definition of the   Helmholtz fluctuation induced  force \cite{DR2022,Dantchev2023b} in the canonical $(T-M)$-ensemble is:
\begin{equation}
	\label{HelmDef}
	\beta F_{\rm H}^{(\zeta)}(L,T,M)\equiv- \frac{\partial}{\partial L}f_{\rm ex}^{(\zeta)}(L,T,M)
\end{equation}
and
\begin{equation}
	\label{excess_free_energy_definition_M}
	f_{\rm ex}^{(\zeta)}(L,T,M) \equiv f^{(\zeta)}(L,T,M)-L f_H(T,m).
\end{equation}
In the above formula, $m=\lim_{L, A\to \infty}M/(LA)$ is the average magnetization, and $f_H(T,m)$  is the Helmholtz free energy density of the ``infinite'' system. In the remainder of this article we will take $L=N a$, where $N$ is an integer number, and without loss of generality  we set the lattice spacing $a=1$.

We stress that the definition and existence of Helmholtz force is by no means limited to the Ising chain and can be addressed, in principle, in any model of interest.

We note that a somewhat elaborate information about  the ensemble behaviour of fluctuation-induced forces has  not  yet been obtained.

In the remainder of the current text,  on the example of the well known one-dimensional Ising model, we  present some new both exact analytical and numerical results for the behavior of the Casimir and Helmholtz forces. We will consider the case of the Ising model with a defect bond.   The definition of the model is given in Sec. \ref{sec:the model}. The derivation of the partition function of the model in GCE is presented in Sec.  \ref{sec:Ising-grand-canonical-ensemble}. General results for the behavior of the free energy density of the finite chain with defect bond and how from the results presented there one easily can obtain those for periodic, antiperiodic and Dirichlet boundary conditions are given in Sec. \ref{sec:bulk-system}. The behavior of the Casimir force in Ising chain with defect bond is discussed in Sec. \ref{sec:Casimir-force-plots}. The derivation of the partition function in canonical ensemble is presented in Sec.  \ref{sec:canonical-partition-function}. The behavior of the Helmholtz is then discussed and visualized in Sec. \ref{sec:Helmholtz-force-plots}. The article closes with a discussion and concluding remarks section \ref{sec:discussion}.

\section{The Model}
\label{sec:the model}
\sectionmark{Ising Model}
 
 Let us consider a one-dimensional Ising chain of $N$ spins $(S_i \pm 1, i=1,.., N)$. We suppose that the interaction between them $J$ is of a ferromagnetic type, i.e., that $J>0$.  The Hamiltonian of the model is given by
\begin{equation}
	\label{eq:Hamiltonian}
{\cal	H}^{(\zeta)}=-J\sum_{i=1}^{N-1} S_i S_{i+1} - J_{BC} S_1 S_N + h\sum_{i=1}^N S_i.
\end{equation}
This form of the Hamiltonian allows for the discussion of different boundary conditions: when $J_{BC}=-J, J, 0, J_a$  one has periodic (PBC's), antiperiodic (ABC's), or Dirichlet-Dirichlet (DBC's) (also termed free, or missing neighbors), boundary conditions, respectively.
Here we focus on the more general case when $J_{BC}=J_a$, where $J_a$ can have both positive or negative values, which we will call a \textit{model with a defect bond}. In the last case we will use the notation $\zeta={\rm db}$.  

The main quantity of interest in the statistical mechanics is the partition function. For the considered model the partition function  is given by
\begin{equation}
Z_{\rm GC}^{\rm (db)}(N,K,K_a, h)=\sum_{\{S_i\}} \exp\left[-\beta  {\cal H}^{{\rm (db)}}\right].
\end{equation}
Here $K=\beta J$ and $K_a=\beta J_a$, while ${\rm (db)}$ depicts the  considered boundary conditions. 

\section{On the Behavior of the Model in Grand Canonical Ensemble}
\label{sec:Ising-grand-canonical-ensemble}
\sectionmark{GCE}

The partition function of the system can be written in the form 
\begin{eqnarray}
	\label{ZIsingH}
Z_{\rm GC}^{\rm (db)}(N,K,K_a, h) &=& \sum_{\{S_i=\pm 1\}} \exp \left[K\left(S_1 S_2+S_2 S_3+\cdots + S_{N-2} S_{N-1} \right) +K_a S_1 S_N\right. \nonumber \\
	&&\left.+h\left(S_1+S_2+\cdots+S_{N-1}+ S_N\right)\right]. 
\end{eqnarray}
It is helpful to cast the above formula as follows 
\begin{eqnarray}
	\label{ZIsingHeq}
&&	Z_{\rm GC}^{\rm db}(N,K,K_a, h)= \\
&& \sum_{\{S_i\}}\exp \left[\frac{1}{2} \left(h S_1+h S_2\right)+K S_1 S_2\right] \exp \left[\frac{1}{2} \left(h S_2+h S_3\right)+K S_2 S_3\right]\nonumber
	 \\
	 && \times \cdots \times \exp \left[\frac{1}{2} \left(h S_{N-1}+h S_N\right)+K S_{N-1} S_N\right] \exp \left[\frac{1}{2} \left(h S_{N}+h S_1\right)+K_a S_{N} S_1\right]. \nonumber
\end{eqnarray}
Introducing the matrices 
\begin{equation}
	\label{eq:matrices_internal_part_chain}
	{\mathbf T}=\begin{pmatrix}
		\exp\left(K+h\right) & \exp\left(-K\right) \\\\
		\exp\left(-K\right) & \exp\left(K-h \right)    
	\end{pmatrix}, \quad \mbox{and} \quad 	{\mathbf T}_a=\begin{pmatrix}
	\exp\left(K_a+h\right) & \exp\left(-K_a\right) \\\\
	\exp\left(-K_a\right) & \exp\left(K_a-h \right)    
	\end{pmatrix},
\end{equation}
it is easy to show that 
\begin{equation}
	\label{eq:Z-via_T}
Z_{\rm GC}^{\rm (db)}(N,K,K_a, h)=	{\rm Tr}\,\left[{\mathbf T}^{N-1}  {\mathbf T}_a\right].
\end{equation}
The matrix ${\mathbf T}$ is usually called transfer matrix.  In the remainder we will need its eigenvalues. They are, customarily, denoted by  $\lambda_{1}$ and $\lambda_2$ and read 
\be
\label{eq:eigenvalues_per1}
\lambda_{1,2}(K, h) = e^K \left(\cosh h \pm \sqrt{e^{-4 K} + \sinh^2 h}\right).
\ee
Obviously, $\lambda_{1},\lambda_2 \in \mathbb{R}$, with $\lambda_{1}>\lambda_2$. The two-dimensional  matrix, which diagonalizes ${\mathbf T}$ is 
\begin{equation}
	\label{eq:orthogonal_matrix_P}
	\mathbf{P}=\begin{pmatrix}
		\cos\phi & -\sin \phi \\ \\
		\sin \phi & \cos \phi  
	\end{pmatrix}, \quad \mbox{with} \quad \mathbf{P}^{-1}\,\mathbf{T}\, \mathbf{P} =\begin{pmatrix}
		\lambda_{1} & 0 \\ \\
		0 & \lambda_2 
	\end{pmatrix},
\end{equation}
where $\phi$ is determined by \cite{B82}
\begin{equation}
	\label{eq:def-phi}
	\cot 2\phi=\exp(2K)\sinh h, \quad 0<\phi<\pi/2.
\end{equation}
Explicitly, one then has
\begin{eqnarray}
	\label{eq:matrix-P-explicit}
	\cos \phi= \frac{1}{\sqrt{2}}\sqrt{1+\frac{\sinh (h)}{\sqrt{\sinh ^2(h)+e^{-4 K}}}}\,, \nonumber\\
	\sin \phi= \frac{1}{\sqrt{2}}\sqrt{1-\frac{\sinh (h)}{\sqrt{\sinh ^2(h)+e^{-4 K}}}}.
\end{eqnarray}
Using the cyclic property of the trace operation we can transform $Z_{\rm GC}^{\rm (db)}(N,K,K_a, h)$ in \eq{eq:Z-via_T} into
\begin{eqnarray}
	\label{eq:diagonalization}
Z_{\rm GC}^{\rm (db)}(N,K,K_a, h) &=& {\rm Tr}\,\left[\left(\mathbf{P}^{-1} {\mathbf T}^{N-1}\mathbf{P}\right)\, \left(\mathbf{P}^{-1} {\mathbf T}_a \, \mathbf{P}\right)\right]\nonumber\\
&=&{\rm Tr}\,\left[\begin{pmatrix}
	\lambda_{1}^{N-1} & 0 \\ \\
	0 & \lambda_2^{N-1} 
\end{pmatrix}\right]\left(\mathbf{P}^{-1} {\mathbf T}_a \, \mathbf{P}\right).
\end{eqnarray}
Performing the calculations in \eq{eq:diagonalization}, we obtain the partition function of the finite Ising chain with a defect bond in the GCE:
\begin{eqnarray}
	\label{eq:Z-final-result}
Z_{\rm GC}^{\rm (db)}(N,K,K_a, h) &=&	\lambda _1^{N-1} \left(e^{K_a} \cosh (h)+ A(K,K_a,h)\right) \nonumber \\
&& +\lambda _2^{N-1} \left(e^{K_a} \cosh (h)-A(K,K_a,h)\right),
\end{eqnarray}
where
\begin{equation}
	\label{eq:A-result}
	A(K,K_a,h)=e^{-K_a} \frac{ \left(\sinh ^2(h) e^{2 (K+K_a)}+1\right)}{\sqrt{e^{4 K} \sinh ^2(h)+1}}.
\end{equation}

% Always give a unique label
% and use \ref{<label>} for cross-references
% and \cite{<label>} for bibliographic references
% use \sectionmark{}
% to alter or adjust the section heading in the running head

\subsection{On the Behavior of the Infinite Ising Chain and Leading  Finite-Size Corrections for the Finite Chain with Defect Bond}
\label{sec:bulk-system} 

The behavior in GCE of the one-dimensional 
Ising model in the thermodynamic limit $N\to \infty$  is discussed in most textbooks on statistical 
mechanics \cite{B82,H87,K2007,PB2011,BH2019} and is well known.  Here we summarize only these of the known results that will be used in the remainder of the current study. 

We first remind that the infinite one-dimensional Ising chain with short-ranged interactions exhibits an essential critical point at $T=0$. About that point the chain demonstrates the usual scaling behavior - see below.  

The free energy density of the "bulk" chain is 
\begin{equation}
	\label{eq:bulk-gibbs-free-energy}
	\beta f_b(K,h)=-\ln \lambda_{1}(K,h),
\end{equation}
with $ \lambda_{1}(K,h)$ given in \eq{eq:eigenvalues_per1}. 
We know the behavior of the correlation length, see, e.g.,  Ref. \cite[p. 36, Eq. 2.2.15]{B82} 
\begin{equation}
	\label{eq:scaling_length_Ising}
	\xi^{-1}(K,h)=\ln \left[\lambda_1(K,h)/\lambda_2(K,h)\right],
\end{equation}
and so one can easily specify the scaling variables.  It is clear that $\xi$ diverges when $\lambda_2\to\lambda_{1}$. Obviously, this happens when $h\to 0$ and $K\to\infty$.  Defining
\be
\label{eq:xit_Ising}
\xi_t\equiv\xi(K,0)\simeq \frac{1}{2}e^{2K}, \;\mbox{when}\; K\gg 1, \mbox{and}\; \xi_h\equiv\lim_{K\to \infty}\xi(K,h)\simeq \frac{1}{2h}, \; \mbox{when} \; h\ll 1,
\ee
for the scaling variables one identifies
\be\label{eq:scaling_variables_Ising}
x_t=N/\xi_t= 2 N e^{-2K}, \quad \mbox{and} \quad x_h =N/\xi_h=2 N h. 
\ee
It immediately follows that the correlation length, the bulk magnetization, and the bulk Gibbs free energy in the limit $K\gg 1$,  in terms of these scaling variables,  read 
\be
\label{eq:corr_length_scaling}
\xi(K,h)=\frac{N}{\sqrt{x_h^2+x_t^2}}, \quad m_b(K,h)=\frac{x_h}{\sqrt{x_h^2+x_t^2}}, \quad \beta f_b(K,h)= -K-\frac{1}{4 N}\sqrt{x_h^2+x_t^2}.
\ee

Next, it must be recalled that in terms of $t=\exp(-2K)$ and $h$, one obtains the usual scaling relations with the scaling exponents, see, e.g., Ref. \cite{B82} 
\be
\label{eq:crit_exponents}
\alpha=\gamma=\nu=\eta=1,\; \beta=0,\;\delta=\infty, \; \mbox{with, however,}\; \beta \delta=1. 
\ee

By definition, the free energy density of the finite chain is,
\begin{equation}
	\label{eq:chain-gibbs-free-energy}
	\beta f^{\rm (db)}(N,K,K_a,h)=-\frac{1}{N} \ln Z^{({\rm db})}_{\rm GC}(N,K,K_a,h).
\end{equation}
From \eq{eq:chain-gibbs-free-energy} and 	\eq{eq:Z-final-result}, one obtains the result 
\begin{equation}
	\label{eq:decomposition-of-free-energy}
\beta f^{\rm (db)}(N,K,K_a,h) = \beta f_b(K,h)+\frac{1}{N}\beta f_{\rm surface}^{\rm (db)}(K,K_a,h)+ \beta \Delta f_N^{\rm (db)}(K,K_a,h),
\end{equation}
where
\begin{equation}
	\label{eq:surface-free-energy}
	\beta f_{\rm surface}^{\rm (db)}(K,K_a,h)=\lambda_1(K,h)-\ln \left[e^{K_a} \cosh (h)+ A(K,K_a,h)\right],
\end{equation}
and 
\begin{equation}
	\label{eq:remnant-part}
	\beta \Delta f_N^{\rm (db)}(K,K_a,h)=-\frac{1}{N}\ln \left\{1+\exp \left[-\frac{N-1}{\xi (K,h)}\right] \;\frac{e^{K_a} \cosh (h)-A(K,K_a,h)}{e^{K_a} \cosh (h)+ A(K,K_a,h)}\right\}.
\end{equation}
Let us briefly discuss the meaning of the terms in 	\eq{eq:decomposition-of-free-energy}.

 First, the term $\beta f_{\rm surface}^{\rm (db)}(K,K_a,h)$ has the  meaning of a "surface free energy density". It shall exist under DBC's but shall vanish for PBC's. One can check that, after some simple algebra, indeed $\beta \Delta f_N^{\rm (db)}(K,K_a=K,h)=0$. For ABC's, when $K_a=-K$ the quantity  $\beta f_{\rm surface}^{\rm (db)}(K,K_a=-K,h)$ has the meaning of an interface free energy, which characterize the interface effected by the imposed ABC's. Apparently, the corresponding result for this case is $\beta f_{\rm surface}^{\rm (db)}(K,K_a=-K,h)=-\ln \left[\cosh (h) \bigg/\sqrt{e^{4 K} \sinh ^2(h)+1}\right]$, which coincide with the one reported in Ref. \cite[Eq. (3.9)]{Dantchev2023b}. 
 
Second, we note that the redundant term $\beta \Delta f_N^{\rm (db)}(K,K_a,h)$ is exponentially small for $N\gg 1$ when $\xi (K,h)={\cal O}(1)$. The only  exception is the case when $N\propto \xi (K,h)$, i.e., $N/\xi (K,h)={\cal O}(1)$. The last relation defines the so-called "finite-size scaling region".  Inherently, it is described by scaling variables $x_t={\cal O}(1)$ and $x_h={\cal O}(1)$, as given in \eq{eq:scaling_variables_Ising}. Thus, the term $\beta \Delta f^{\rm (db)}(K,K_a,h)$ cannot  be neglected and, as we will see below, is of a primarily importance for the behavior of the Casimir force.

\section{Behavior of the Casimir Force in Ising Chain with Defect Bond}
\label{sec:Casimir-force-plots}

In accordance with the definitions Eqs. \eqref{CasDef}, \eqref{excess_free_energy_definition}, from 
 Eqs. 	\eqref{eq:decomposition-of-free-energy} -- 	\eqref{eq:remnant-part} for the Casimir force we obtain the following result
\begin{equation}
	\label{eq:Casimir-nonzero-field}
	\beta F_{\rm Cas}^{\rm (db)}(N,K,h)=-\frac{1}{N} \frac{N}{ \xi (K,h)} \frac{1}{r(K,K_a,h) \exp \left[(N-1)/\xi (K,h)\right]+1}, 
\end{equation}
where 
\begin{equation}
	\label{eq:r-definition}
	r[K,K_a,h]:=\frac{e^{K_a} \cosh (h)+A(K,K_a,h)}{e^{K_a} \cosh (h)- A(K,K_a,h)}.
\end{equation}
\begin{figure}[htb]
	\sidecaption[t]
	% For example, with the option graphics use
	\includegraphics[scale=.35]{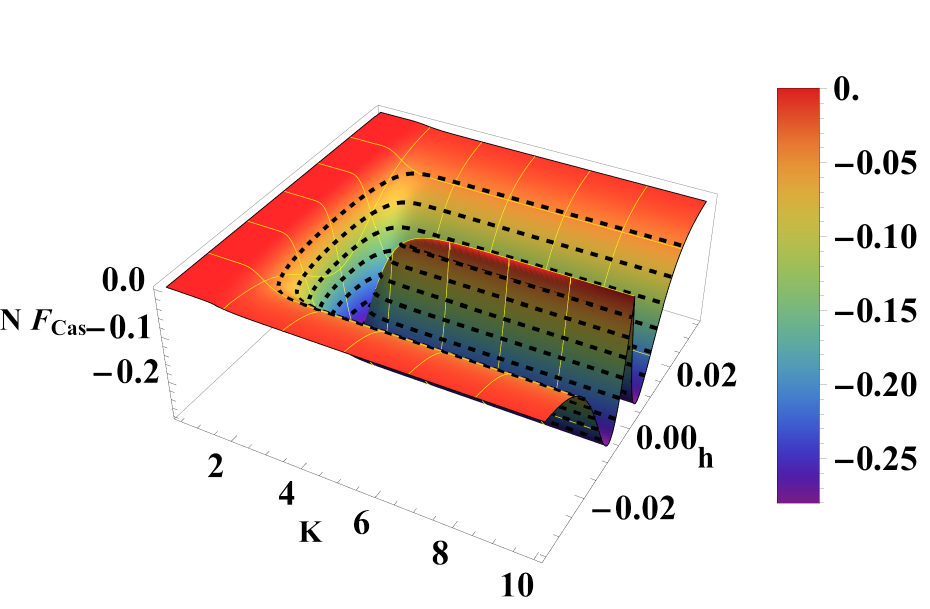} \quad \includegraphics[scale=.35]{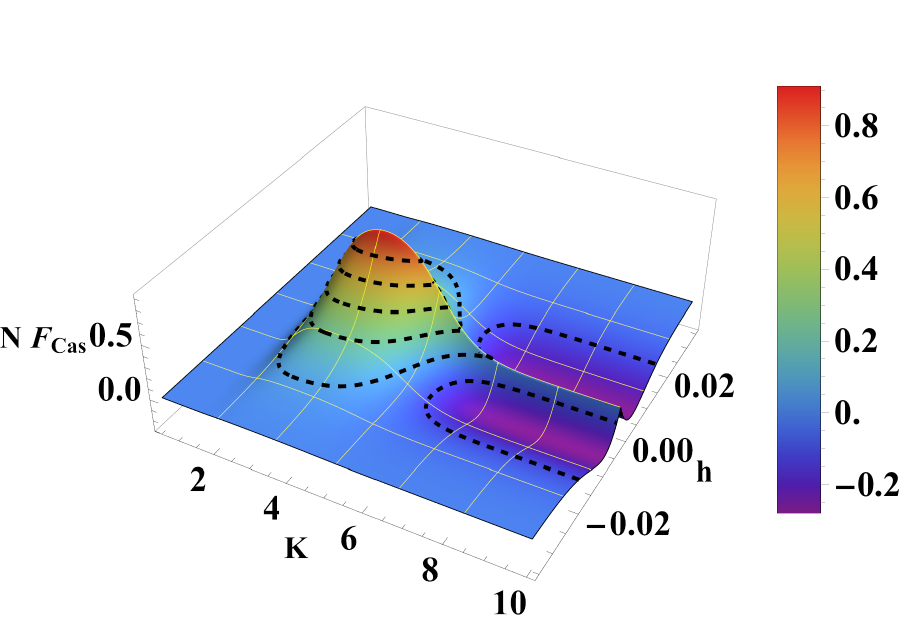} 
	\caption{If $K_a>0$ the Casimir force is \textit{attractive} (the corresponding function is negative). In the opposite case - with $K_a<0$, the force can be both \textit{attractive and repulsive}, depending on $K$ and $h$. The force is always symmetric with respect of the sign of the external magnetic field $h$. In the current figure the \textbf{left panel} shows the behavior of the Casimir force for $N=100$ with $K_a=3$. The \textbf{right panel} depicts  the behavior of the Casimir for with again  $N=100$ but for $K_a=-3$. The behavior of the force is shown as a function of $K$ and $h$.}
	\label{fig:Casimir-3D}     
\end{figure}
The behavior of the Casimir force as a function of $K$, $K_a$ and $h$ for $N=100$ and $K_a=\pm 3$ is shown in Fig. 	\ref{fig:Casimir-3D}.     

Let us now consider the behavior of the force in the scaling regime. On general grounds we expect that the scaling function of the Casimir force $X_{\rm Cas}(x_t,x_a,x_h)$ is 
\begin{equation}
	\label{eq:scaling-function-Casimir-definition}
	\beta F_{\rm Cas}^{\rm (db)}(N,K,h)=\frac{1}{N} X_{\rm Cas}^{\rm (db)}(x_t,K_a,x_h).
\end{equation}
 From \eq{eq:scaling_variables_Ising}, and Eqs. \eqref{eq:Casimir-nonzero-field} -- \eqref{eq:scaling-function-Casimir-definition} one derives the corresponding explicit expressions 
\begin{equation}
	\label{eq:scaling-function-Casimir}
	X_{\rm Cas}^{\rm (db)}(x_t,K_a,x_h)=-\frac{\sqrt{x_h^2+x_t^2}}{r(x_t,K_a,x_h) \exp \left(\sqrt{x_h^2+x_t^2}\right) +1}, 
\end{equation}
where
\begin{equation}
	\label{eq:scaling-Casimir}
	r(x_t,K_a,x_h)= \frac{\sqrt{x_h^2+x_t^2}+x_t \exp(-2K_a)}{\sqrt{x_h^2+x_t^2}-x_t\exp(-2 K_a)}.
\end{equation}
Obviously, if  $r(x_t,K_a,x_h)>0$ one has $X_{\rm Cas}(x_t,K_a,x_h)<0$. Furthermore,  $X_{\rm Cas}(x_t,K_a,x_h)$ decays exponentially when $x_h^2+x_t^2\gg 1$.

\begin{figure}[h!]
	\centering
	\includegraphics[width=2.2in]{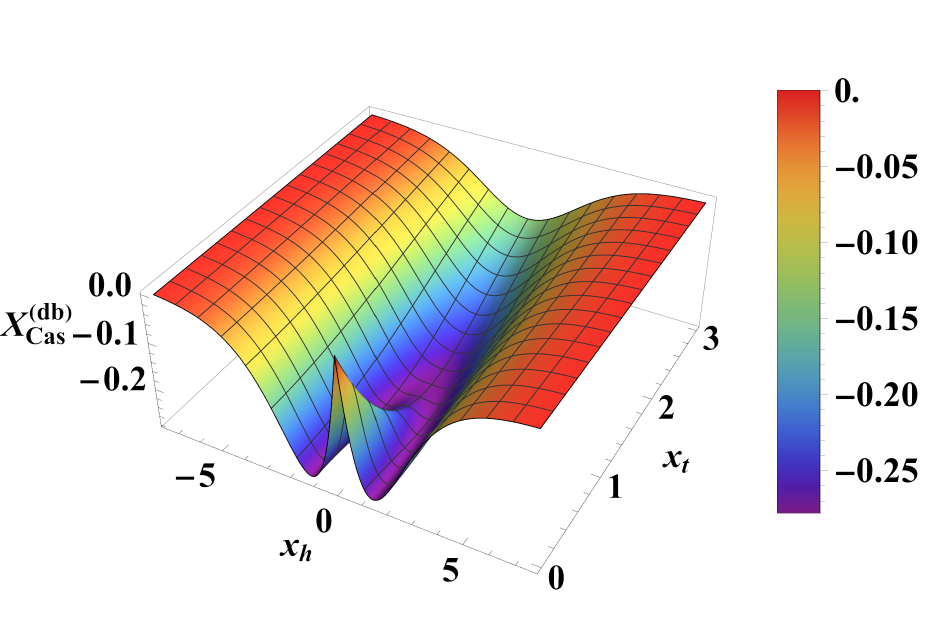} \quad
	\includegraphics[width=2.2in]{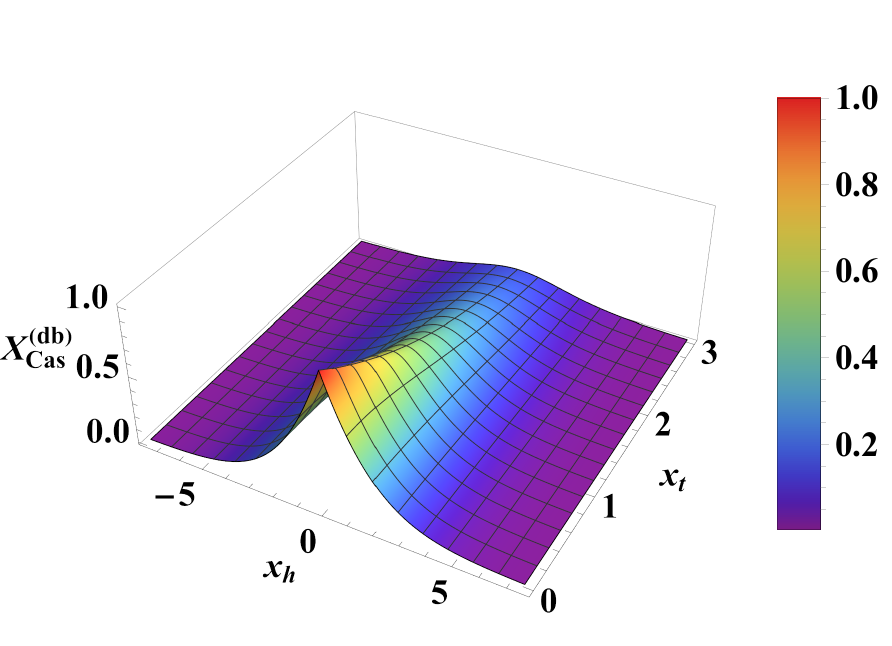}
	\caption{On the \textbf{left panel}: The behavior of the  scaling function $X_{\rm Cas}^{\rm (db)}(x_t,K_a,x_h)$  of the Casimir force as a function of the scaling variables $x_t$ and $x_h$ for $K_a=3$. We observe that the function is negative  for all values of $x_t$ and $x_h$, i.e., the force is \textit{attractive}. On the\textbf{ right panel}: The behavior of the function $X_{\rm Cas}^{\rm (db)}(x_t,K_a,x_h)$ for $K_a=-3$. We observe that the force is \textit{repulsive}.}
	\label{fig:3D-Casimir-scaling}
\end{figure}

The behavior of the scaling function $X_{\rm Cas}^{\rm (db)}(x_t,K_a,x_h)$  of the Casimir force are visualized in Fig. \ref{fig:3D-Casimir-scaling}. We observe, that the force is  symmetric with respect to the sign of $x_h $, as  must be the case. 

\section{On the Behavior of the Model in Canonical Ensemble}
\label{sec:canonical-partition-function}
In the canonical ensemble the total magnetization $M$ of the chain is fixed. This constrain can be expressed by using an integral presentation of the Kronecker delta-function  
\begin{equation}
	\delta[S,M]=
	\frac{1}{2\pi}\int_{-\pi}^{+\pi}e^{i(S-M)\phi}d\phi,\quad S=\sum_{i=1} ^N S_{i},\quad S,M \in \mathbb{Z}.
\end{equation}
Then the canonical partition function  is given by
\begin{equation}
	Z^{\rm (db)}_C(N,K,K_a,M)=\sum_{\{S_i\}_{\rm (db)}}
	e^{-\beta {\cal H}} \delta[S,M],\quad  
\end{equation}
where ${\cal H}$ is given by \eq{eq:Hamiltonian} with $J_{BC}=J_a$.
Here the symbol  $\{S_i\}_{\rm (db)}$ means that the set of spins obeys the boundary conditions with a defect bond.
Further we have
\begin{eqnarray}
	\label{eq:transfer-matrix-db}
	Z^{(\rm db)}_C(N,K,K_a,M) &=&\frac{1}{2\pi}\int_{-\pi}^{\pi}{e^{-i M \phi}\left[\sum_{\{S_i\}^{\rm (db)}} e^{-\beta {\cal H}+i \phi \sum_{i} S_i}\right]} d\phi \nonumber
	\\ 
	&=&\frac{1}{2\pi}\int_{-\pi}^{\pi}{e^{-i M \phi}} \;Z^{(\rm db)}_{\rm GC}(N,K,K_a,i \phi)\,d\phi.
\end{eqnarray}
Using \eq{eq:Z-final-result}, as well as \cite[see there Eqs. (2.9) and (2.11)]{Dantchev2024}, one derives
\begin{eqnarray}
	\label{GKDab}
&& Z^{(\rm db)}_{\rm GC}(N,K,K_a,i \phi)
	=2\left[\sqrt{2\sinh(2K)}
	\right]^{N-1} \\
&&	\times \bigg\{e^{K_a}\cos(\phi) T_{N-1}\bigg(\tilde {z}(K,\phi\bigg)
	+ \frac{e^{-(K_a+K)}-e^{K+K_a}\sinh^2(\phi)}{\sqrt{2\sinh(2K)}} U_{N-2}
	\bigg(\tilde{z}(K,\phi)\bigg)\bigg\}, \nonumber
\end{eqnarray} 
where 
\begin{equation}
	\label{Nsvo}
	\tilde{z}(K,\phi)\equiv z(K,i\phi)=	 \frac{e^K\cos(\phi)}{\sqrt{2\sinh(2K)}},\quad K>0,\quad \phi \in [-\pi, \pi].
\end{equation}
Plugging 	\eq{GKDab} into \eq{eq:transfer-matrix-db} after using the properties of the Chebyshev's polynomials \cite[8.941.1-8.941.3]{GR}  it can be shown that 
\begin{eqnarray}
	\label {AAA}
	Z^{\rm (db)}_C(N,K,K_a,M)&=&\left[\sqrt{2\sinh(2K)}
		\right]^{N}\bigg\{	\bigg[e^{K_a-K}-\frac{\sinh(K_a+K)}{\sinh(2K)}\bigg]  {\rm D}\left(N,M;e^{-4K}\right)\nonumber \\
		&& + \frac{1}{2}e^{K-K_a}	\bigg[\frac{e^{2K_a}-e^{-2K}}{\sinh(2K)} \bigg]{\rm I}(N,M,e^{-4K})\bigg\}.
\end{eqnarray}	
Here
\begin{equation}
	\label{ppp-new}
	{\rm I}(N,M,z) :=\frac{4}{\pi}\int_0^{\pi/2}\cos(M x)\;
	T_{N}\left(\frac{\cos(x)}{\sqrt{1-z}}\right)dx,
\end{equation}
and
\begin{eqnarray}
	\label{babx-new}
	{\rm D}\left(N,M;z\right)=
	\frac{4}{\pi}\int_{0}^{\pi/2}\cos(M x)  \frac{\cos(x)}{\sqrt{1-z}}U_{N-1}\left(\frac{\cos(x)}{\sqrt{1-z}}\right)\,dx.
\end{eqnarray}
As shown in Ref. \cite{Dantchev2023b} and \cite{Dantchev2024}, the above integrals can be expressed in terms of the Gauss hypergeometric functions. The results are 
\begin{equation}
	\label{eq:I-expression}
	{\rm I}(N,M,z)= N z (1-z)^{-N/2}\; _2F_1\left(\frac{1}{2}(M-N+2),\frac{1}{2}(-M-N+2);2,z\right),
\end{equation}
and 
\begin{eqnarray}
	\label{eq:D-expression}
	&& {\rm D}(N,M,z)=(1-z)^{-N/2}z\; 
	\bigg\{N\; _2F_1\left(\frac{1}{2}(M-N+2),\frac{1}{2}(-M-N+2);2,z\right)+\nonumber\\
	&&2(z^{-1}-1)\;_2F_1\left(\frac{1}{2}(M-N+2),\frac{1}{2}(-M-N+2);1,z\right)\bigg\}. 
\end{eqnarray}
Thus, we have derived in an exact explicit form the partition function of the one-dimensional Ising model with fixed magnetization $M$ possessing a defect bond $K_a$. Now we pass to its scaling behavior. Using the asymptotic expansion \cite[Eq. (D.6)]{Dantchev2023b}, which in the current notations reads
\begin{align} 
	\label{eq:expansion-D6}	
	_2F_1(\frac{1}{2}(Nm-N+2),&\frac{1}{2}(-Nm-N+2);\gamma;e^{-4K})\simeq \\
	& (\gamma-1)! \left(\frac{1}{4} \sqrt{1-m^2}\, x_t\right)^{1-\gamma}
	\left[I_{\gamma-1}\left( \frac{1}{2}\sqrt{1-m^2}\, x_t\right)
	+O(N^{-1})\right], \nonumber
\end{align}
from \eq{AAA}, and Eqs. \eqref{eq:I-expression}, \eqref{eq:D-expression}, we obtain
\begin{eqnarray}
	\label{MF}
e^{-KN} && Z^{(D)}_C(N,K,K_a,M) = \\
&&	{\cal A}(K,K_a)\frac{I_{1}\left( \frac{1}{2}\sqrt{1-m^2}\, x_t\right)}{\sqrt{1-m^2}}+{\cal B}(K,K_a)I_{0}\left(\frac{1}{2} \sqrt{1-m^2}\, x_t\right), \nonumber
\end{eqnarray}
where, for $K \gg 1$:
\begin{equation}
%{\cal A}(K,K_a)
	{\cal A}=2e^{-2K} \bigg\{\bigg[e^{K_a-K}-\frac{\sinh(K_a+K)}{\sinh(2K)}\bigg]  
	+ \frac{1}{2}
	e^{-K_a}	\bigg[\frac{e^{2K_a}-e^{-2K}}{e^{-K}\sinh(2K)} \bigg]\bigg\} \underset{K\gg 1}{\simeq}   2 e^{K_a-3K}.  
\end{equation}
and
\begin{eqnarray}
	{\cal B}=2\bigg[e^{K_a-K}-\frac{\sinh(K_a+K)}
{\sinh(2K)}\bigg]\underset{K\gg 1}{\simeq} 2 e^{-K_a-3K}.
\end{eqnarray}

\section{Behavior of the Helmholtz Force in Ising Chain with Defect Bond}
\label{sec:Helmholtz-force-plots}

\begin{figure}[htb]
	\sidecaption[t]
	% For example, with the option graphics use
\includegraphics[scale=.35]{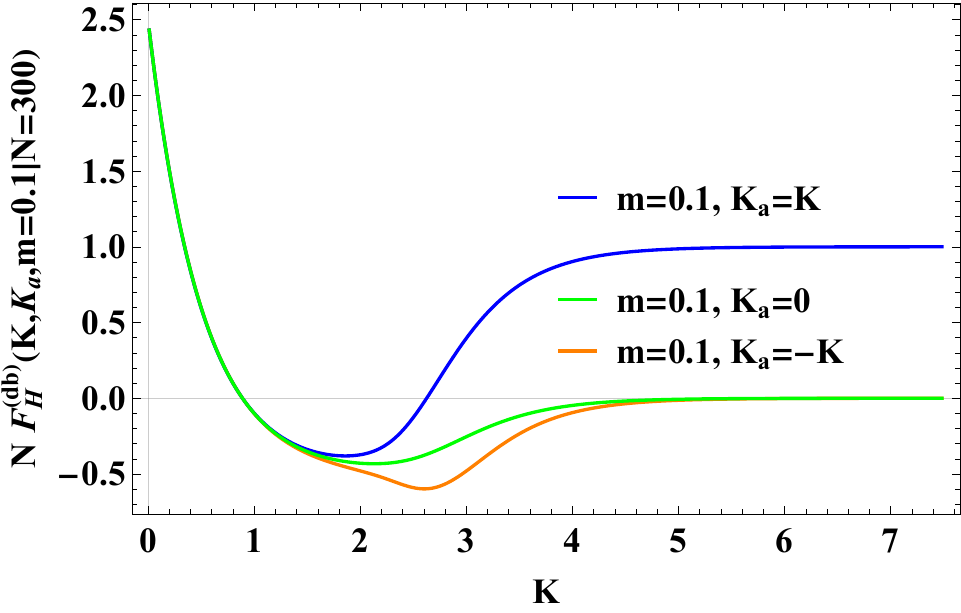} \quad	\includegraphics[scale=.32]{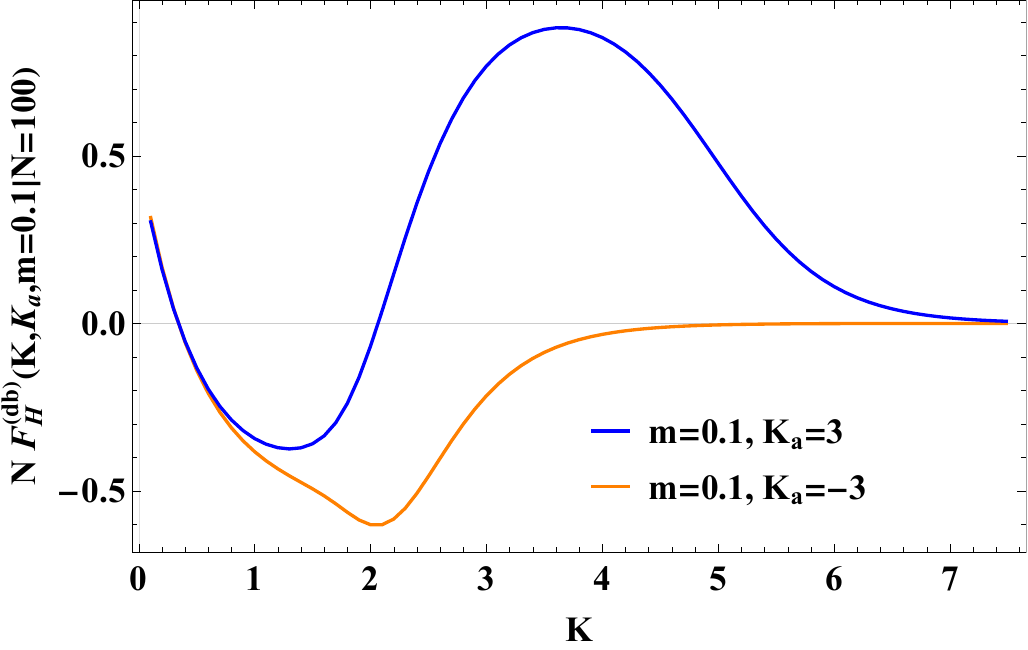}
	\caption{If $K_a>0$ the Helmholtz force is similar to the corresponding result for PBC's - see Refs. \cite{DR2022,Dantchev2023b,Dantchev2024}. In the opposite case - with $K_a<0$, the force resembles the one for ABC's - see Refs. \cite{Dantchev2023b,Dantchev2024}.  The force is always symmetric with respect of the sign of $M$. In the current figure the \textbf{left panel} shows the behavior of the Helmholtz force for $N=100$ with $K_a=\pm K$. The \textbf{right panel} depicts  the behavior of the force  with again  $N=100$ but for $K_a=\pm 3$.}
	\label{fig:Helmholtz-force-m01}     
\end{figure}

\begin{figure}[htb]
	\sidecaption[t]
	% For example, with the option graphics use
\includegraphics[scale=.3]{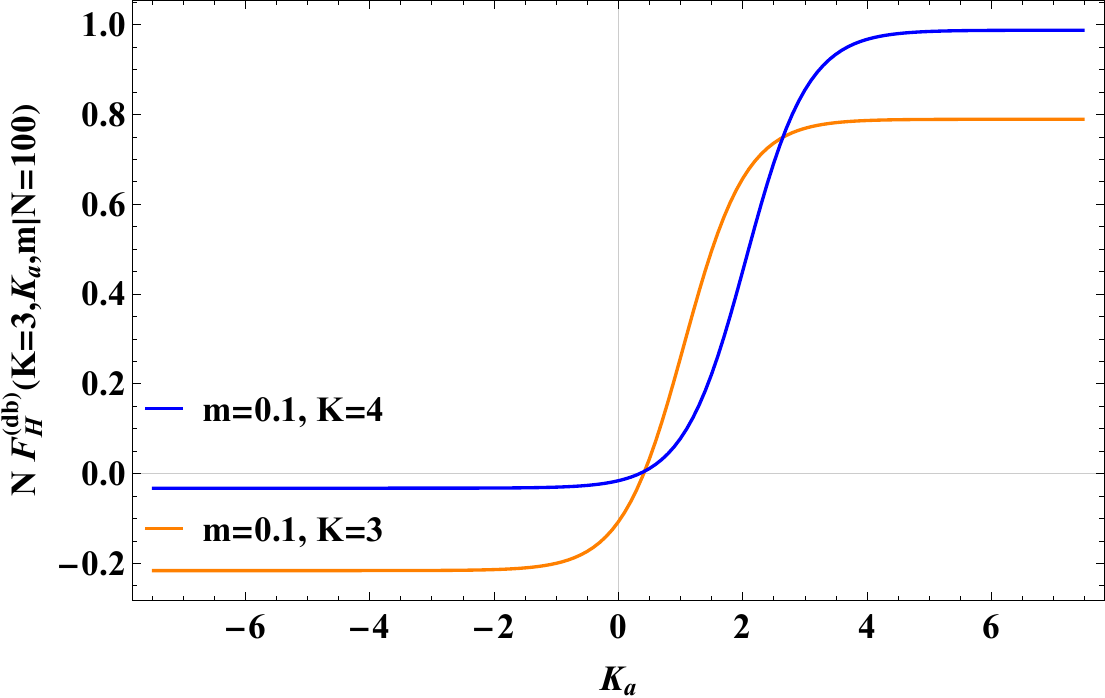} \quad	
\includegraphics[scale=.32]{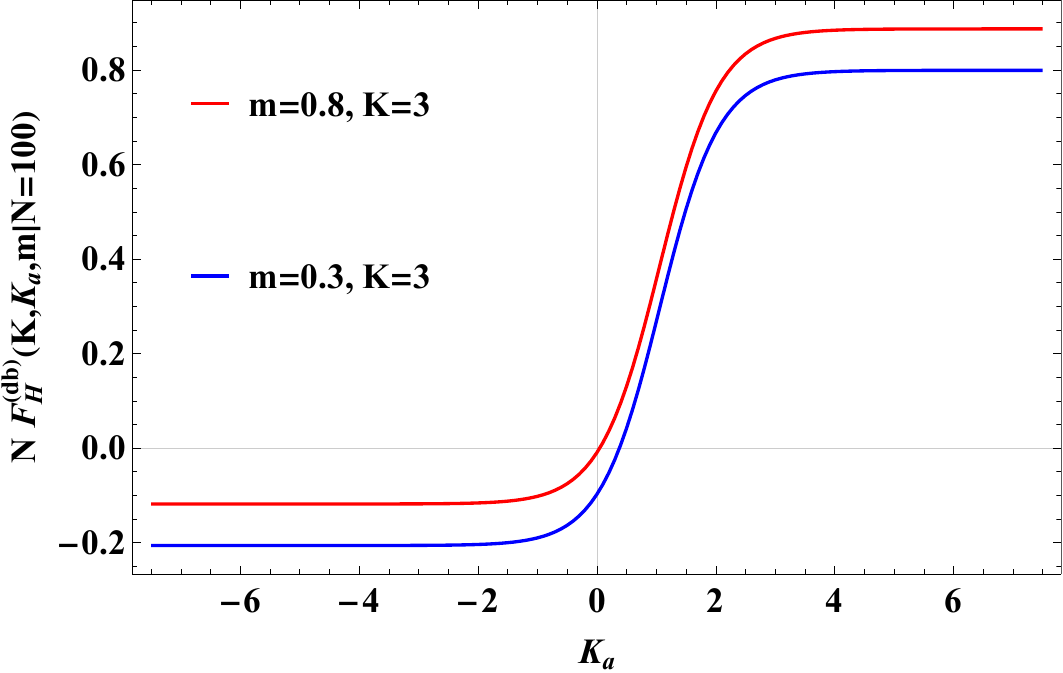} 
	\caption{The Helmholtz force as a function of $K_a$ for two fixed pairs of values of $m$ and $K$. The \textbf{left panel}  demonstrates the influence of $K$ for a fixed value of $m=0.1$, while the right panel shows the role of $m$ for a fixed value of $K=3$. We observe that for large values of $K$ the force is very small for negative values of $K_a$. For moderate values of $K$ (say $K=3$) the force is a negative constant for negative values of $K_a$, and positive for large values of $K_a$.}
	\label{fig:Helmholtz-force-as-a-function-of-Ka}     
\end{figure}

Based on Eqs. \eqref{AAA} -- \eqref{eq:D-expression} we are ready to derive the behavior of the Helmholtz force defined in Eqs. 	\eqref{HelmDef} -- \eqref{excess_free_energy_definition_M}. The only additional information we still need is the behavior of the bulk Helmholtz free energy density. It is, see Ref. \cite{Dantchev2023b} and Ref. \cite{Dantchev2024}
\begin{align}
	\label{eq:Helmholtz_free_energy_bulk_explicit}
	\beta a_b(K,m) &= K + \frac{1}{2}\ln(1 - m^2) \\
	&- \ln\left[1 + \sqrt{m^2 + e^{4 K} (1 - m^2)}\right]  + \; m \sinh ^{-1}\left(\frac{e^{-2 K} m}{\sqrt{1-m^2}}\right). \nonumber
\end{align}
The behavior of the Helmholtz force is shown in Figs. \ref{fig:Helmholtz-force-m01} and \ref{fig:Helmholtz-force-as-a-function-of-Ka} . In     Fig. \ref{fig:Helmholtz-force-m01} the force is visualized as a function of $K$. The left panel of the figure shows the behavior of the force for magnetization $m=0.1$ and for the three limiting cases of the values of the coupling constant: $K_a=K$, when our system is equivalent to the one with periodic boundary conditions, for $K_a=-K$ when it represents a system under antiperiodic boundary conditions, and with $K_a=0$ when it turns into a system with Dirichlet boundary conditions. We see that the obtained curves, calculated for $N=300$, agree completely with the ones reported in Refs. \cite{DR2022,Dantchev2023b,Dantchev2024}. The right panel of Fig. \ref{fig:Helmholtz-force-m01} shows the behavior of the Helmholtz force for $N=100$ with the values of $K_a$ fixed ate $K_a=\pm 3$. We observe that for moderate values of $K$ the behavior of the force essentially differ in the two sub-case having, however, the same asymptotic for large and small values of $K$. 

The Helmholtz force as a function of $K_a$ for two pairs of fixed values of $m$ and $K$ is depicted in Fig. \ref{fig:Helmholtz-force-as-a-function-of-Ka}    as a function of $K_a$. The left panel demonstrates the influence of $K$ when $m$ is fixed, while the right panel show the complimentary case - the role of the value of $m$ when $K$ is fixed. 

\section{Discussion and concluding remarks}
\label{sec:discussion}

In the current article we have presented a brief review of some of the fluctuation induced forces. This have been done in Sec. \ref{sec:introduction}. We commented on the QED and critical Casimir force, as well as on the newly introduced Helmholtz force.  Some theoretical questions of practical application in nano- and micro-world have been outlined and discussed. Then the behavior of the critical Casimir and Helmholtz forces have been considered on the example of the  one-dimensional Ising model in grand canonical and in the canonical statistical mechanical ensembles. The model is with one defect bond and is defined in Sec. \ref{sec:the model}. 

The behavior of the model in grand canonical ensemble is studied in Sec. \ref{sec:Ising-grand-canonical-ensemble}. The main result there is the exact expression for the partition function of the model given in \eq{eq:Z-final-result} and 	\eq{eq:A-result}.  

The behavior of the Casimir force is investigated in Sec. \ref{sec:Casimir-force-plots}. The main results are visualized in Figs. \ref{fig:Casimir-3D}   and 	\ref{fig:3D-Casimir-scaling}. 
\begin{itemize}
	\item Fig. \ref{fig:Casimir-3D} shows the behavior of the force for general values of the basic two parameters of the model - the strength of the coupling constant $K$ and the external magnetic field $h$. The calculations are performed for $N=100$. We observe that when the defect bond $K_a$ is of a ferromagnetic type, see the left panel, force is attractive and symmetric, as expected, with respect to the sign of the external magnetic field. 
	\item When the defect bond is of antiferromagnetic type. i.e., $K_a<0$ the behavior of the force is much more interesting in that it can be both attractive \textit{and} repulsive - see the right panel of the figure. 
	\item On the left panel of Fig. \ref{fig:3D-Casimir-scaling} it is shown the scaling function of the Casimir force $X_{\rm Cas}^{\rm (db)}(x_t,K_a=3,x_h)$, with the scaling variables $x_t$ and $x_h$ defined in 
	\eq{eq:scaling_variables_Ising}. Since $K_a>0$ the behavior of the force is similar to that of the periodic boundary, conditions, i.e., the force is \textit{attractive}. 
	
	\item  In  the right panel of Fig. \ref{fig:3D-Casimir-scaling} the opposite case of  $X_{\rm Cas}^{\rm (db)}(x_t,K_a=-3,x_h)$. As we see, with $K_a<0$ the force resembles the one for antiperiodic boundary conditions and is always \textit{repulsive}. 
\end{itemize}

The behavior of the model in the  canonical ensemble is considered in Sec. \ref{sec:canonical-partition-function}. Again, the main result there is the explicit exact expression for the partition function of the model given in Eqs.\eqref{AAA} -- 	\eqref{eq:D-expression}. It is presented there in terms of the Gauss hypergeometric functions. 

The behavior of the Helmholtz Force in Ising chain with defect bond is considered in Sec. 
\ref{sec:Helmholtz-force-plots}. The basic results are depicted in Figs. \ref{fig:Helmholtz-force-m01} and \ref{fig:Helmholtz-force-as-a-function-of-Ka}.     
\begin{itemize}
	\item Fig. \ref{fig:Helmholtz-force-m01} represents the behavior of the Helmholtz force  as a function of the coupling constant $K$. On the left panel few principal cases of the value of $K_a$, namely $K_a=K, K_a=-K$ and $K_a=0$ are shown. The value of the magnetization is fixed to $m=0.1$. We observe that the force can be \textit{both} attractive \textit{and} repulsive and coincides with the behavior of the system with periodic, antiperiodic and Dirichlet, boundary conditions, respectively. 
	\item The right panel of Fig. \ref{fig:Helmholtz-force-m01} clearly shows that the precise value of $K_a$, with all other parameters kept the same, is important for the behavior of the force. 
	\item Fig. \ref{fig:Helmholtz-force-as-a-function-of-Ka} depicts the  Helmholtz force as a function of $K_a$. We observe that for large values of $K$ the force is very small for negative values of $K_a$. The force changes sign for moderate values of $K$ (say $K=3$): it is attractive for negative values of $K_a$, and repulsive for large values of $K_a$.
	
\end{itemize}

We find that all significant results are consistent with the expectations of finite-size scaling theory \cite{BDT2000}. 

The present article demonstrates that the behavior of the fluctuation induced forces crucially depend on the statistical ensemble in which they are defined and also on the presence of impurity in the system. These important issues have not been intensively studied yet. 

\begin{acknowledgement}
The partial financial support via Grant No KP-06-H72/5 of Bulgarian NSF is gratefully acknowledged.
\end{acknowledgement}
\ethics{Competing Interests}{
This study was funded by the Bulgarian NSF via Grant No KP-06-H72/5.  \newline
The authors have no conflicts of interest to declare that are relevant to the content of this chapter.}

%\eject

\ethics{Ethics Approval}{\newline 
 Informed consent to publish was obtained from the individual participants of the article.}


\begin{thebibliography}{10}
	\providecommand{\url}[1]{{#1}}
	\providecommand{\urlprefix}{URL }
	\expandafter\ifx\csname urlstyle\endcsname\relax
	\providecommand{\doi}[1]{DOI \discretionary{}{}{}#1}\else
	\providecommand{\doi}{DOI \discretionary{}{}{}\begingroup
		\urlstyle{rm}\Url}\fi
	
	\bibitem{DR2022}
	D.~Dantchev, J.~Rudnick, Phys. Rev. E \textbf{106}, L042103 (2022).
	\newblock \doi{10.1103/PhysRevE.106.L042103}.
	\newblock \urlprefix\url{https://link.aps.org/doi/10.1103/PhysRevE.106.L042103}
	
	\bibitem{DD2022}
	D.~Dantchev, S.~Dietrich, Physics Reports \textbf{1005}, 1 (2023).
	\newblock \doi{https://doi.org/10.1016/j.physrep.2022.12.004}.
	\newblock
	\urlprefix\url{https://www.sciencedirect.com/science/article/pii/S0370157322004070}
	
	\bibitem{E07}
	A.~Einstein, Annalen der Physik \textbf{327}(3), 569 (1907).
	\newblock \doi{10.1002/andp.19073270311}.
	\newblock \urlprefix\url{http://dx.doi.org/10.1002/andp.19073270311}
	
	\bibitem{C48}
	H.B. Casimir, Proc. K. Ned. Akad. Wet. \textbf{51}, 793 (1948)
	
	\bibitem{MT97}
	V.M. Mostepanenko, N.N. Trunov, \emph{The Casimir effect and its applications}
	(Energoatomizdat, Moscow, 1990, in Russian; English version: Clarendon, New
	York, 1997)
	
	\bibitem{KG99}
	M.~Kardar, R.~Golestanian, Rev. Mod. Phys. \textbf{71}(4), 1233 (1999).
	\newblock \doi{10.1103/RevModPhys.71.1233}
	
	\bibitem{M2001}
	K.A. Milton, \emph{The Casimir Effect: Physical Manifestations of Zero-point
		Energy} (World Scientific, Singapore, 2001)
	
	\bibitem{BKMM2009}
	M.~Bordag, G.L. Klimchitskaya, U.~Mohideen, V.M. Mostepanenko, \emph{Advances
		in the Casimir effect} (Oxford University Press, Oxford, 2009)
	
	\bibitem{M2004}
	K.A. Milton, J. Phys. A: Math. Gen. \textbf{37}, R209 (2004)
	
	\bibitem{GLR2008}
	C.~Genet, A.~Lambrecht, S.~Reynaud, Eur. Phys. J. Special Topics \textbf{160},
	183–193 (2008)
	
	\bibitem{KMM2011}
	G.L. Klimchitskaya, U.~Mohideen, V.M. Mostepanenko, Int. J. Mod. Phys. B
	\textbf{25}, 171 (2011).
	\newblock \doi{10.1142/S0217979211057736}
	
	\bibitem{RCJ2011}
	A.W. Rodriguez, F.~Capasso, S.G. Johnson, Nature Photonics \textbf{5},
	211–221 (2011).
	\newblock \doi{doi:10.1038/nphoton.2011.39}
	
	\bibitem{FAKA2014}
	A.~Farrokhabadi, N.~Abadian, F.~Kanjouri, M.~Abadyan, Int. J. Mod. Phys. B
	\textbf{28}(19), 1450129 (2014).
	\newblock \doi{10.1142/S021797921450129X}.
	\newblock
	\urlprefix\url{http://www.worldscientific.com/doi/abs/10.1142/S021797921450129X}
	
	\bibitem{FMRA2014}
	A.~Farrokhabadi, J.~Mokhtari, R.~Rach, M.~Abadyan, Int. J. Mod. Phys. B
	\textbf{29}(02), 1450245 (2015).
	\newblock \doi{10.1142/S0217979214502452}.
	\newblock
	\urlprefix\url{http://www.worldscientific.com/doi/abs/10.1142/S0217979214502452}
	
	\bibitem{L56}
	E.M. Lifshitz, Sov. Phys. \textbf{2}, 73; (1956).
	\newblock {Zhur. Eksptl. i Teoret. Fiz.}{\bf 29}, 94--110 (1955) (In Russian)
	
	\bibitem{L.E.Dzyaloshinskii1961}
	E.M.L. L.~E.~Dzyaloshinskii, L.P. Pitaevskii, Adv. Phys. \textbf{10}, 165
	(1961)
	
	\bibitem{MN76}
	J.~Mahanty, B.W. Ninham, \emph{Dispersion Forces} (Academic, New York, 1976)
	
	\bibitem{P2006}
	V.A. Parsegian, \emph{Van der Waals Forces} (Cambridge University Press, New
	York, 2006)
	
	\bibitem{DLP61r}
	I.E. Dzyaloshinskii, E.M. Lifshitz, L.P. Pitaevskii, Sov. Phys. Usp.
	\textbf{4}, 153  (1961).
	\newblock In Russian: Usp. Fiz. Nauk {\bf 73}, 381 (1961).
	
	\bibitem{Di88}
	S.~Dietrich, in \emph{Phase Transitions and Critical Phenomena}, vol.~12, ed.
	by C.~Domb, J.L. Lebowitz (Academic, New York, 1988), pp. 1--218
	
	\bibitem{MCP2009}
	J.N. Munday, F.~Capasso, V.A. Parsegian, Nature \textbf{457}(7226), 170 (2009).
	\newblock \urlprefix\url{http://dx.doi.org/10.1038/nature07610}
	
	\bibitem{FG78}
	M.E. Fisher, P.G. de~Gennes, C. R. Seances Acad. Sci. Paris Ser. B
	\textbf{287}, 207 (1978)
	
	\bibitem{Krech1994}
	M.~Krech, \emph{The Casimir Effect in Critical Systems} (World Scientific,
	Singapore, 1994)
	
	\bibitem{BDT2000}
	J.G. Brankov, D.M. Dantchev, N.S. Tonchev, \emph{The Theory of Critical
		Phenomena in Finite-Size Systems - Scaling and Quantum Effects} (World
	Scientific, Singapore, 2000)
	
	\bibitem{MD2018}
	A.~Macio\l{}ek, S.~Dietrich, Rev. Mod. Phys. \textbf{90}, 045001 (2018).
	\newblock \doi{10.1103/RevModPhys.90.045001}.
	\newblock \urlprefix\url{https://link.aps.org/doi/10.1103/RevModPhys.90.045001}
	
	\bibitem{Gambassi2023}
	A.~Gambassi, S.~Dietrich,  \doi{10.48550/ARXIV.2312.15482}
	
	\bibitem{Dantchev2023b}
	D.M. Dantchev, N.S. Tonchev, J.~Rudnick, Annals of Physics \textbf{459},
	169533.
	\newblock \doi{https://doi.org/10.1016/j.aop.2023.169533}.
	\newblock
	\urlprefix\url{https://www.sciencedirect.com/science/article/pii/S0003491623003354}
	
	\bibitem{B82}
	R.J. Baxter, \emph{Exactly Solved Models in Statistical Mechanics} (Academic,
	London, 1982)
	
	\bibitem{H87}
	K.~Huang, \emph{Statistical Mechanics}, 2nd edn. (John Wiley \& Sons, New York,
	1987)
	
	\bibitem{K2007}
	M.~Kardar, \emph{Statistical Physics of Fields} (Cambridge University Press,
	Cambridge, 2007)
	
	\bibitem{PB2011}
	R.K. Pathria, P.D. Beale, \emph{Statistical Mechanics}, 3rd edn. (Elsevier,
	Amsterdam, 2011)
	
	\bibitem{BH2019}
	B.~Liebchen, H.~Löwen, The Journal of Chemical Physics \textbf{150}(6), 061102
	(2019).
	\newblock \doi{10.1063/1.5082284}.
	\newblock \urlprefix\url{https://doi.org/10.1063/1.5082284}
	
	\bibitem{Dantchev2024}
	D.M. Dantchev, N.S. Tonchev, J.~Rudnick,  \doi{10.48550/ARXIV.2402.04459}
	
	\bibitem{GR}
	I.S. Gradshteyn, I.H. Ryzhik, \emph{Table of Integrals, Series, and Products}
	(Academic, New York, 2007)
	
\end{thebibliography}
\end{document}